\documentclass[letterpaper, 10 pt, conference]{ieeeconf}  

\newcommand{\fullonly}[1]{#1} 
\newcommand{\shortonly}[1]{} 
\newcommand{\hiddenonly}[1]{}

\IEEEoverridecommandlockouts                              

\usepackage{mathtools}
\DeclarePairedDelimiter\ceil{\lceil}{\rceil}

\usepackage[style=ieee]{biblatex} 
\usepackage[percent]{overpic}

\usepackage{graphicx}
\usepackage[dvipsnames]{xcolor}

\usepackage{scalerel}
\usepackage{tikz}
\usetikzlibrary{svg.path}

\definecolor{orcidlogocol}{HTML}{A6CE39}
\tikzset{
    orcidlogo/.pic={
        \fill[orcidlogocol] svg{M256,128c0,70.7-57.3,128-128,128C57.3,256,0,198.7,0,128C0,57.3,57.3,0,128,0C198.7,0,256,57.3,256,128z};
        \fill[white] svg{M86.3,186.2H70.9V79.1h15.4v48.4V186.2z}
        svg{M108.9,79.1h41.6c39.6,0,57,28.3,57,53.6c0,27.5-21.5,53.6-56.8,53.6h-41.8V79.1z M124.3,172.4h24.5c34.9,0,42.9-26.5,42.9-39.7c0-21.5-13.7-39.7-43.7-39.7h-23.7V172.4z}
        svg{M88.7,56.8c0,5.5-4.5,10.1-10.1,10.1c-5.6,0-10.1-4.6-10.1-10.1c0-5.6,4.5-10.1,10.1-10.1C84.2,46.7,88.7,51.3,88.7,56.8z};
    }
}

\newcommand\orcidicon[1]{\href{https://orcid.org/#1}{\mbox{\scalerel*{
                \begin{tikzpicture}[yscale=-1,transform shape]
                \pic{orcidlogo};
                \end{tikzpicture}
            }{|}}}}

\usepackage{subcaption}
\makeatletter
\renewcommand*\subcaption@label{%
  \caption@withoptargs\subcaption@@label}
\makeatother
\usepackage[all=normal,wordspacing=tight,bibbreaks=tight,paragraphs=tight,floats=tight,lists=tight,charwidths=tight]{savetrees}
\usepackage{verbatim}

\usepackage{amsmath}
\DeclareMathOperator*{\argmin}{argmin} 
\newcommand\norm[1]{\left\lVert#1\right\rVert}
\usepackage{esvect}
\usepackage{booktabs}
\usepackage{adjustbox}
\usepackage{array}
\usepackage{multirow}

\newcolumntype{R}[2]{%
    >{\adjustbox{angle=#1,lap=\width-(#2)}\bgroup}%
    l%
    <{\egroup}%
}

\newcommand{\todo}[1]{\textcolor{red}{\textbf{TODO:} #1}}

\title{\LARGE \bf
Multi-Agent Spatial Predictive Control\\with Application to Drone Flocking (Extended Version)
}

\author{Andreas Brandstätter$^{1a}$, Scott A. Smolka$^{2}$, Scott D. Stoller$^{2}$, Ashish Tiwari$^{3}$, Radu Grosu$^{1}$
\thanks{$^{1}$ CPS, Technische Universit\"at Wien (TU Wien), Austria}%
\thanks{$^{2}$ Department of Computer Science, Stony Brook University, USA}%
\thanks{$^{3}$ Microsoft, USA}%
\thanks{$^{a}$ Correspondence: {\tt\scriptsize andreas.brandstaetter@tuwien.ac.at}}%
}

\begin{document}


\maketitle
\thispagestyle{empty}
\pagestyle{empty}

\begin{abstract}


We introduce the novel concept of \emph{Spatial Predictive Control} (SPC) to solve the following problem:
given a collection of agents (e.g., drones) with positional low-level controllers (LLCs) and a mission-specific distributed cost function, how can a distributed controller achieve and maintain cost-function minimization without a plant model and only positional observations of the environment?
Our fully distributed SPC controller
is based strictly on the position of the agent itself and on those of its neighboring agents. This information is used in every time-step to compute the gradient of the cost function and to perform a spatial look-ahead to predict the best next target position for the LLC.
Using a high-fidelity simulation environment, we show that SPC outperforms the most closely related class of controllers, Potential Field Controllers, on the drone flocking problem.
We also show that SPC is able to cope with a potential sim-to-real transfer gap by demonstrating its performance on real hardware, namely our implementation of flocking using nine Crazyflie~2.1 drones.
\end{abstract}

\section{INTRODUCTION}



\noindent Drone swarms, a quintessential example of a multi-agent system, can carry out tasks that cannot be accomplished by individual drones alone~\cite{survey}. They can, for example, collectively carry a heavy load while still being much more agile than a single larger drone~\cite{michael2011cooperative,TransportationVison}. In search-and-rescue applications, a swarm can explore unknown terrain by covering individual paths that jointly cover the entire area~\cite{dronefleet,dronerecharge,Michael-2012-17115}. 
These collective maneuvers can be expressed as the problem of minimizing a {\em positional cost function}, i.e., a cost function that depends on the positions of the drones (and possibly information about their environment).

Off-the-shelf drones, such as Crazyflie~\cite{Crazyflie}, DJI~\cite{dji-missions}, and Parrot~\cite{parrot-autonomous}, come equipped with a \emph{positional low-level controller} (LLC). Such a controller takes a position argument as input and maneuvers the drone to this position, where it then hovers. Unfortunately, their code is often proprietary and the exact parameters of the physical drone model might not be available.
Positional LLCs are common in other types of robots as well, including the Landshark~\cite{landshark} and Taurob~\cite{taurob} unmanned ground vehicles, and the Bluefin\textsuperscript{\textregistered}-12~\cite{bluefin} unmanned underwater vehicle.
%
%
%
%

\begin{figure}[t]
    \centering
    \includegraphics[width=0.99\columnwidth]{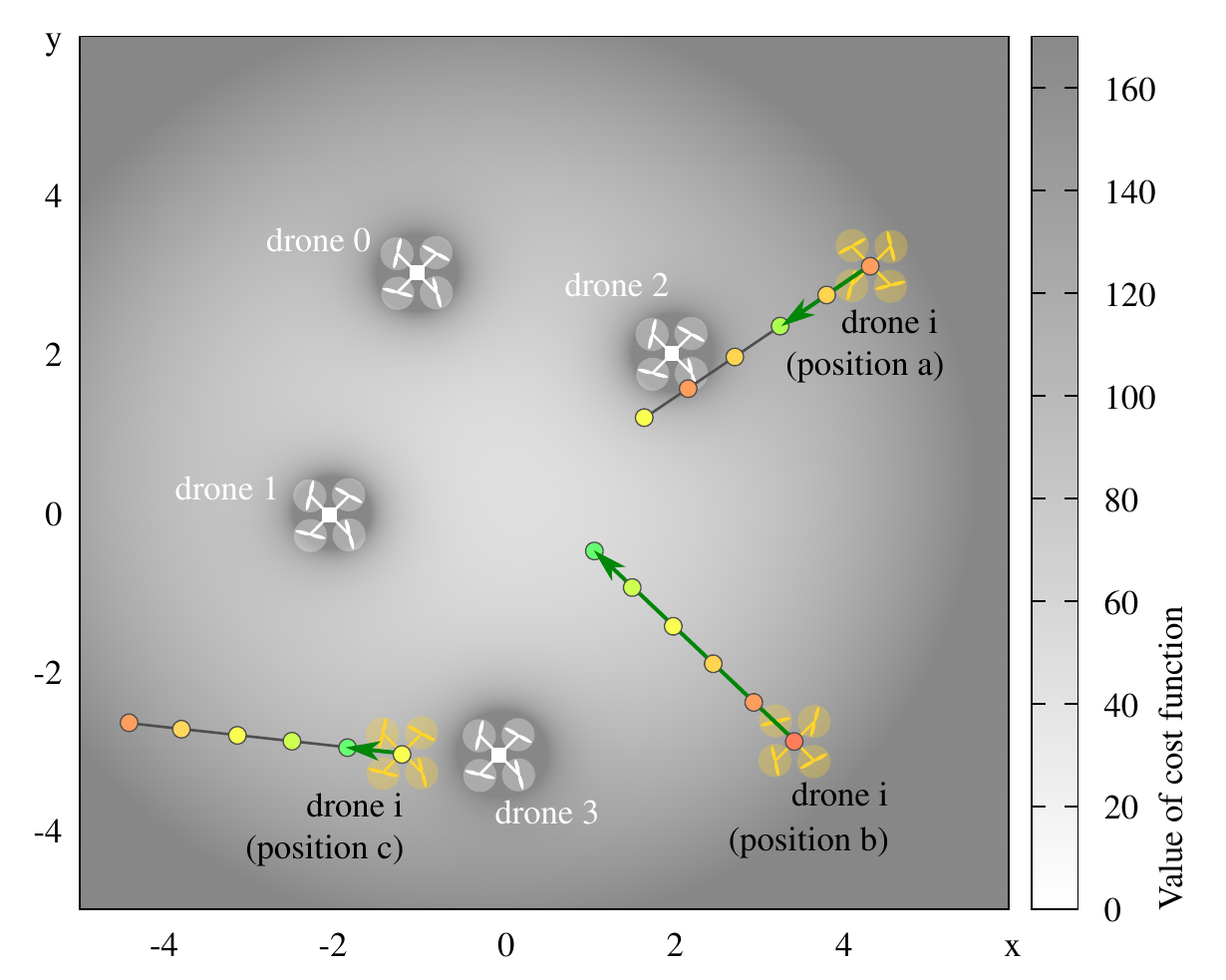}
    \caption{For a given set of neighboring drones (drones~0 to~3 in white), the cost function for drone $i$ is shown as a grayscale heatmap.  The direction in which drone $i$ (in yellow) should move is determined by the gradient of the cost function (for positions~\emph{a}, \emph{b}, and~\emph{c}). SPC evaluates the cost at the spatial lookahead (indicated by colored dots) along this direction and chooses the best value for drone $i$'s next position (green arrow).
    }
    \label{fig:gradient}
\end{figure}

In this paper, we address the following \textbf{Challenge Problem:}
\emph{Design a distributed controller that minimizes a given positional cost function
for robots with positional LLCs, no available model of the plant dynamics, and only positional observations of their environment.}

To solve this problem, we introduce \emph{Spatial Predictive Control} (SPC), a novel distributed high-level approach to multi-agent control.
In SPC, each agent's controller identifies $N$ equally-spaced points within a maximum look-ahead distance $\epsilon \cdot N$ from 
its current position and in the direction of the negative gradient of a cost function $c$. The SPC controller then computes the value of the cost function for each of these points and chooses the one with minimal cost as the target location to be sent to the positional LLC.

\textit{SPC vs Planning.}
Given that agents (robots) are equipped with LLCs, one might ask if a controller is actually needed, or would a planning-based approach suffice.
Based on the initial positions of agents and obstacles, a plan of way-points could be generated for the LLC to follow. However, since the environment is constantly changing due to the movement of other agents (and possibly obstacles), such a plan would become outdated very quickly.
This is exactly what we consider in our SPC: in every time step, we use the observations of the environment to calculate the next input to the LLC; the LLC makes a best effort to reach this position. 
Hence, the key difference between SPC and planning is the granularity of the time horizon: planning uses long time horizons for calculating trajectories, whereas in our approach, feedback from the plant in every time step is used to recalculate the desired next position.

\textit{SPC vs MPC.}
To solve the challenge problem, one might consider the LLC as part of the plant and design a Model Predictive Control (MPC) for the high-level control. Such an approach is not directly applicable since neither a dynamic model of the physical plant nor the internals of the LLC are available. Even if an approximate dynamic model could be obtained using system identification techniques, and if the code for the LLC is available (e.g., for Crazyflies), MPC remains a computationally expensive method~\cite{negenborn}. This is especially relevant for embedded processors with limited computing capabilities.
In contrast, SPC does not require a plant model and avoids extensive prediction calculations.

\textit{SPC vs PFC.}
Potential Field Controllers (PFCs) are well-known controllers for mobile robots. Prior work \cite{Tanner03stabilityof,schwager2009gradient} has considered their application to flocking.  In these approaches, however, agent accelerations are computed, which are not directly applicable to positional LLCs. 
More importantly, PFC takes into account only the gradient of the cost function at the current position. If the cost function is nonlinear as in Figure~\ref{fig:gradient} (e.g., has peaks), then using only the gradient at the current position to determine the next control input may lead to reduced performance.  In contrast, SPC evaluates the cost function at multiple candidate future positions within the spatial look-ahead horizon 
to find the best next position, based on cost-function minimization.

\textbf{Application to Drone Flocking:}
After introducing SPC as a general approach, we apply it to a distributed multi-agent system with the goal of achieving flock formation, and maintaining flock formation while moving to a specified target location, avoiding obstacles in the process.



The cost function $c$
for this problem
(see Section~\ref{sec:costfunction})
depends only on the locations of the drones. As illustrated for four drones in Figure~\ref{fig:gradient}, the (negation of) the gradient of $c$ (see Section~\ref{sec:gradient}) suggests a direction of movement for the drone.  The SPC algorithm (see Section~\ref{sec:spc}) evaluates its local cost function at multiple points in that direction. As the figure illustrates, this helps the drone see peaks and valleys in the cost function that may lie ahead,
even when locally the direction looks promising. 

In summary, the \textbf{main contributions of this paper} are the following.
\begin{itemize}
\item
We introduce the novel concept of \emph{Spatial Predictive Control}, a control methodology
well suited for positional LLCs. 
\item
SPC is \emph{model-free}.  One only needs to be able to determine the cost at different locations along the direction of the cost function's gradient in order to apply it.  Also, 
SPC does not need to measure the velocity or acceleration of neighboring drones.
\item
We evaluate SPC using drone flocking in a high-fidelity drone simulation environment.
\item
We further experimentally validate our approach by achieving flocking with real drone hardware in the form of off-the-shelf Crazyflie quadcopters.
\end{itemize}

\fullonly{
\noindent
\textbf{Paper Outline:} 
Section~\ref{sec:spc} introduces the control problem and presents SPC.
Section~\ref{sec:flocking} introduces our flocking cost function
and related performance metrics.
Section~\ref{sec:eval} presents the results of our experimental evaluation.  Section~\ref{sec:related} considers related work.  Section~\ref{sec:conclusion} offers our concluding remarks.
}

\section{SPC for multi-agent systems} 
\label{sec:spc}

We describe the distributed control problem addressed in this paper and then present SPC for solving that problem.

\subsection{Distributed Control for Distributed Cost in Presence of positional LLCs}
We consider a multi-agent system consisting of a
set $D$ of agents. 
Every agent $i\in D$ has a state $(x_{io},x_{ih})$, where
$x_{io}$ is the observable part of its state and 
$x_{ih}$ is the hidden part of its state.
Agent $i$ has a control input $u_i$ and its dynamics is
assumed to be given by some unknown function $f$:
 $$
   (\frac{d{x_{io}(t)}}{dt},\frac{dx_{ih}(t)}{dt}) = f(t, x_{io}(t), x_{ih}(t), u_i(t))
 $$

Agent $i$ has access to the observable state
of a subset $H_i \subseteq D$ of agents.
$H_i$ will be referred to as the \emph{neighborhood} of $i$.

The objective
for the multi-agent system is given
in terms of a cost function
$c(x_{io}, x_{Hio})$ that 
maps the observable states
of an agent ($x_{io}$) and its neighbors ($x_{Hio}$) to a non-negative real value.
Here we use $x_{Hio}$ as shorthand for $(x_{jo})_{j\in H_i}$.
The goal is for each agent $i$ to minimize $c(x_{io},x_{Hio})$.

In our setting, we do not have ability to directly set $u_i$.
Instead, we can only set a \emph{reference value} $x_{io}^{(r)}$
that is then used by some black-box 
low-level controller
$LLC$ to internally set the control input.
$$
 u_i(t) = LLC(t, x_{io}(t), x_{ih}(t), x_{io}^{(r)})
$$

Both the dynamics of each agent (the function $f$) and the details of the LLC (the function $LLC$) are unknown. The cost function $c$ is given.
We want to find a procedure that allows each agent to minimize its cost in the above setting.

\subsection{Spatial Predictive Control (SPC)}

Let $\nabla_{x_{io}}c(x_{io},x_{Hio})$ denote the gradient of cost function $c$ with respect to $x_{io}$.
One way to minimize the cost $c(x_{io}, x_{Hio})$ would be
to follow the negative of the gradient at every point.
However, if the cost function is nonlinear (e.g., has peaks), then gradients can be misleading. Our key observation is that each agent should \emph{look ahead in the observable state space}, in the direction of the negative gradient, to determine the reference value for its observable state.
The controller picks $N$ equally-spaced points within a maximum look-ahead distance $\epsilon \cdot N$ from the current observable state and in the direction of the negative gradient, 
where $\epsilon$ and $N$ are parameters of the controller.
%
At any given state $(x_{io}(t), x_{Hio}(t))$,
the set $Q_i$ containing these equally-spaced points is given by: 
\begin{equation}
    Q_i = \left\{ x_{io}(t) - n \cdot \epsilon \cdot \frac{\nabla_{x_{io}} c{(x_{io}(t),x_{Hio}(t))}}
    {\norm{\nabla_{x_{io}} c{(x_{io}(t),x_{Hio}(t)}}} \mid n\!=\! 1\!\dots\!N \right\}
\label{eq:grad:spat01}
\end{equation}
Here 
    ${\nabla_{x_{io}} c{(x_{io}(t),x_{Hio}(t)}}$
    denote the evaluation of the gradient of $c$ at the point
    $(x_{io}(t),x_{Hio}(t))$.
Our spatial-predictive controller selects the point in $Q_i$ with minimum cost as the next
target position $x_{io}^{(r)}$ for agent $i$:
\begin{equation}
    x_{io}^{(r)} = \argmin_{\tilde{x}_{io} \in Q_i} \left( c(\tilde{x}_{io}, x_{Hio}(t)) \right)
\label{eq:grad:spat02}
\end{equation}

Note that the spatial predictive controller recomputes the
reference $x_{io}^{(r)}$ at each time step. 
This is important because this computation of the
reference ignores the motion of the neighbors.
However, SPC will respond to any change
in neighbors' observable states in its next computation
of the reference.

\section{Application to Flocking}
\label{sec:flocking}

This section starts with background on flocking, 
describes 
how to use SPC for this application, and then presents metrics to assess the quality of a flocking controller.

\subsection{What is a Flock?}
A set of agents $D$ is in a flock formation if the distance between every pair of agents is ``not too large and not too small.''
These requirements yield the first two terms of our cost function: the \emph{cohesion} and \emph{separation} terms.
These two terms are sufficient to cause the agents to form and maintain a flock formation.


%
In our model, drone $i$ has access to positions of only a subset $H_i$ of drones, namely its local neighbors.  Hence, we define a local cost function, parameterized by $i$, which uses only the positions of drones in $\{i\}\cup H_i$.

\subsection{Cost Function}
\label{sec:costfunction}

\begin{figure}[t]
    \centering
     \begin{overpic}[width=0.85\columnwidth]{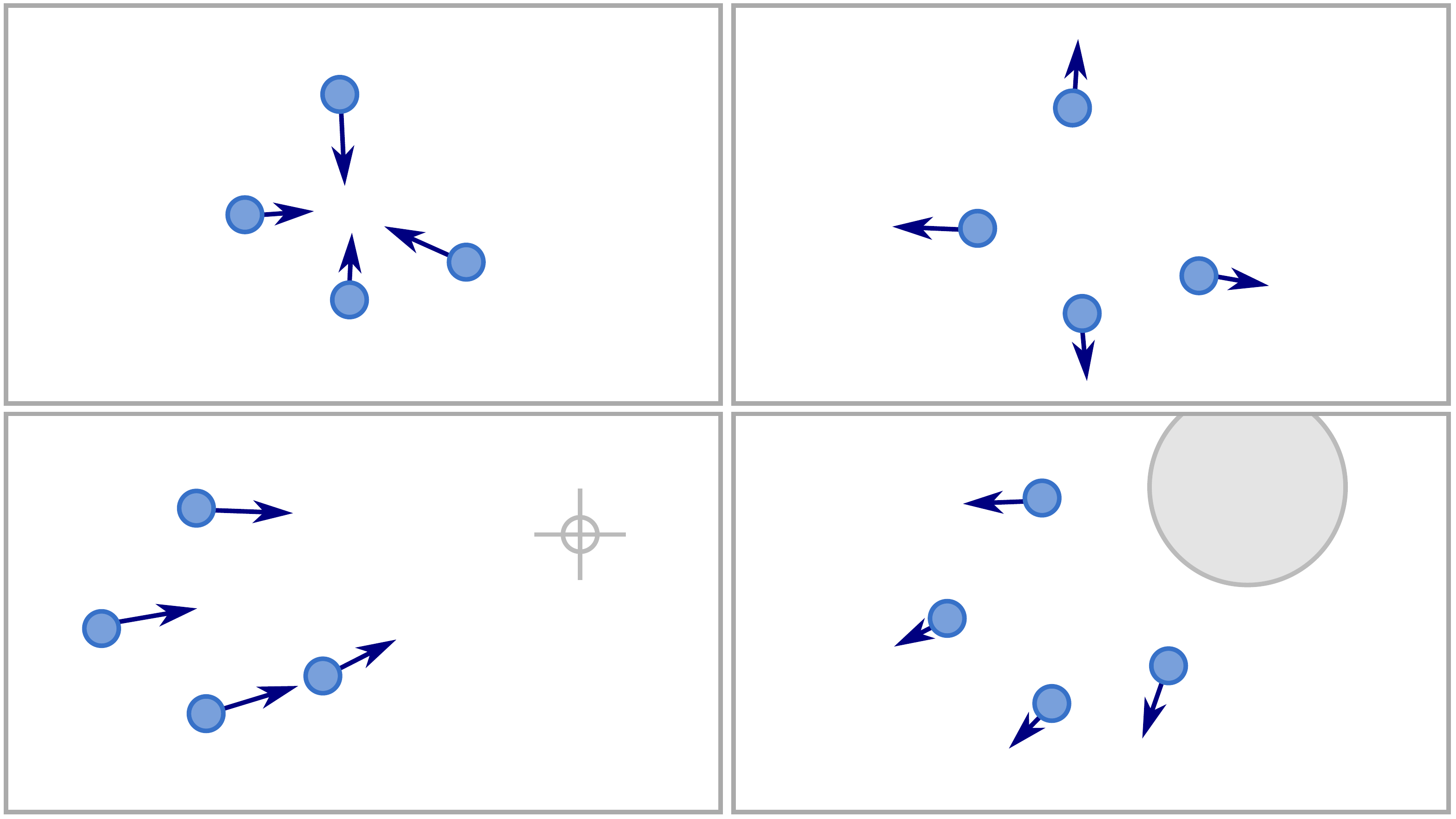}
        \put (2,51) {\textbf{a)}}
        \put (52,51) {\textbf{b)}}
        \put (2,23) {\textbf{c)}}
        \put (52,23) {\textbf{d)}}
     \end{overpic}
    \caption{Directional movements (indicated by arrows) induced by cost-function terms: \textbf{a}:~\emph{Cohesion},  \textbf{b}:~\emph{separation}, \textbf{c}:~\emph{target seeking}, and \textbf{d}:~\emph{obstacle avoidance}.}
    \label{fig:costterms}
\end{figure}
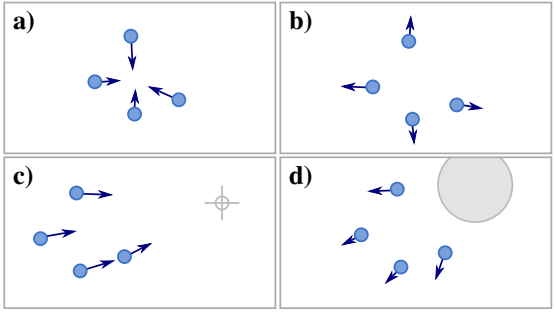

Consider a drone $i$, where $i$ in $D$.
Let $p_j$, when it appears in the local cost function of drone $i$, denote the position of drone $j$ as known to drone $i$; this may differ from the actual position due to sensing error.
Let $p_{H_i}$ denote the tuple of positions of drones in $H_i$ and $r_d$ the radius of each drone.
We define the cost function $c(p_i, p_{H_i})$ as:
%
\begin{align}
    c(p_i,p_{H_i}) = &c_{coh}(p_i,p_{H_i}) + c_{sep}(p_i,p_{H_i}) +  \nonumber
    \\ & c_{tar}(p_i,p_{H_i}) + c_{obs}(p_i)
\label{eq:costFunction}
\end{align}
%

The value of the \emph{cohesion} term increases as drones drift apart, and the \emph{separation} term increases as drones get closer together.  Each term has a weight, denoted by a subscripted $\omega$. \\[1mm]
\noindent\emph{Cohesion term}:
\begin{equation}
    c_{coh}(p_i,p_{H_i}) = \omega_{coh} \cdot \frac{1}{|H_i|} \cdot \sum_{j \in H_i} { \norm{p_i - p_j} }^2
\end{equation}
\noindent\emph{Separation term}:
\begin{equation}
    c_{sep}(p_i,p_{H_i}) = \omega_{sep} \cdot \frac{1}{|H_i|} \cdot \sum_{j \in H_i} \frac{1}{{ max(\norm{p_i\,{-}\,p_j} - 2 r_d,\hat{0}) }^2}
\end{equation}
\noindent The function $max(.,\hat{0})$ ensures positive values when there is sensor noise, but does not further influence the cost function; $\hat{0}$ denotes a very small positive value.

The mission-specific \emph{target seeking} term sets a target location, denoted by $p_{tar}$, for the entire flock.  The \emph{obstacle avoidance} term prevents the drones from colliding with infinitely tall
cylindrical objects of radius $r_k$.  Let $K$ denote the set of obstacles, and $p_k$ denote the $k^{\mathrm{th}}$ obstacle's  center on the xy-plane.\\[1mm] 
\noindent\emph{Target-seeking term}:
\begin{equation}
    c_{tar}(p_i,p_{H_i}) = \omega_{tar} \cdot \norm{ p_{tar} - \frac{p_i + \sum_{j \in H_i}{p_j}} {|H_i| + 1} } ^2
\end{equation}

\noindent\emph{Obstacle-avoidance term}: 
\begin{equation}
c_{obs}(p_i) = \omega_{obs} \cdot \frac{1}{|k|} \cdot \sum_{k \in K} \frac{1}{{ max(\norm{\mathcal{P}(p_i) - p_k}\!-\!r_k\!-\!r_d,\hat{0}) }^2}
\end{equation}
\noindent The function $\mathcal{P}(.)$ projects a vector to the xy-plane.

\subsection{Gradient of the cost function}
\label{sec:gradient}

For SPC, the gradient of the cost function is required in Equation \eqref{eq:grad:spat01}. The gradient is
(for readability, we elide function arguments):
\begin{equation}
\nabla_{p_i}c = \nabla_{p_i}c_{coh} + \nabla_{p_i}c_{sep} + \nabla_{p_i}c_{tar} + \nabla_{p_i}c_{obs}
\end{equation}

\noindent\emph{Cohesion gradient}:
\fullonly{
\begin{equation}
\nabla_{p_i} c_{coh} = \omega_{coh} \cdot \frac{1}{|H_i|} \cdot \sum_{j \in H_i} \nabla_{p_i} { \norm{p_j - p_i} }^2
\label{eq:grad:coh}
\vspace*{-4mm}
\end{equation}
Hence:
}
\begin{equation}
\nabla_{p_i} c_{coh} = 2 \cdot \omega_{coh} \cdot \left( p_i - \frac{1}{|H_i|} \cdot \sum_{j \in H_i} p_j \right)
\label{eq:grad:coh}
\end{equation}

\noindent\emph{Separation gradient}:
\fullonly{
\begin{equation}
\nabla_{p_i} c_{sep} = \omega_{sep} \cdot \frac{1}{|H_i|} \cdot \sum_{j \in H_i} \nabla_{p_i} {{ \norm{p_i - p_j} }^{-2}}
\label{eq:grad:sep}
\vspace*{-4mm}
\end{equation}
\hspace*{8mm}Hence:\vspace*{-2mm}
}
\begin{equation}
\hspace*{-8mm}\nabla_{p_i} c_{sep}\,{=}\,\frac{2 \cdot \omega_{sep}}{|H_i|}\,{\cdot}\!\sum_{j \in H_i} 
{
\frac{p_j - p_i}
{ {(\norm{p_i\!-\!p_j}\!-\!2r_d) }^3\!\cdot\!\norm{p_i\!-\!p_j} }
}
\label{eq:grad:sep}
\end{equation}

\noindent\emph{Target-seeking gradient}:
\fullonly{
\begin{equation}
\nabla_{p_i} c_{tar} = \omega_{tar} \cdot \nabla_{p_i} \norm{ p_{tar} - \frac{p_i + \sum_{j \in H_i}{p_j}} {|H_i| + 1} }^2
\vspace*{-2mm}
\end{equation}
\hspace*{6mm}Hence:\vspace*{-2mm}
}
\begin{equation}
\nabla_{p_i} c_{tar} = \frac{2 \cdot \omega_{tar}} {|H_i| + 1} \cdot 
\left( \frac{p_i + \sum_{j \in H_i}{p_j}} {|H_i| + 1} - p_{tar} \right)
\end{equation}

\noindent\emph{Obstacle-avoidance gradient}:
\fullonly{
($\norm{p_i\,{-}\,p_k} \,{>}],r_k$ should hold)
\begin{equation}
\nabla_{p_i} c_{obs} = \omega_{obs} \cdot \frac{1}{|k|} \cdot \sum_{k \in K} \nabla_{p_i}  \frac{1}{{ (\norm{p_i - p_k} - r_k) }^2}
\end{equation}
\hspace*{3mm}Hence:\vspace*{-2mm}
}
\begin{equation}
\nabla_{p_i} c_{obs}\,{=}\,\frac{2\,{\cdot}\,\omega_{obs}}{|k|}\,{\cdot}\!\sum_{k \in K} \frac{p_k\,{-}\,\mathcal{P}(p_i)}{{ (\norm{\mathcal{P}(p_i)\!-\!p_k}\!-\!r_k\!-\!r_d) }^3 \cdot \norm{\mathcal{P}(p_i)\,{-}\,p_k}} 
\label{eq:grad:obs2}
\end{equation}

\subsection{Flock-Formation Quality metrics}
\label{sec:qualitymetrics}

We define three quality metrics to assess the quality of the flock formation achieved by a flocking controller.\\[1mm]
\noindent\emph{Collision avoidance}:
To avoid collisions, the distance between all pairs of drones must remain above a specified threshold $dist_{thr}$.  We define a metric for the minimum distance between any pair of drones as follows:
\begin{equation}
dist_{min} = \min_{i, j \in D; i \neq j} \norm{p_i - p_j}
\end{equation}
We set $dist_{thr} = 2 \cdot r_{\mathit{drone}} + r_{\mathit{safety}}$, where $r_{\mathit{drone}}$ is the radius of the drone, and $r_{\mathit{safety}}$ is a safety margin.\\[1mm]
\noindent\emph{Compactness}:
Compactness of the flock is measured by the maximum distance of any drone from the centroid of the flock. It is defined as follows:
\begin{equation}
comp_{max} = \max_{i \in D} \norm{\frac{\sum_{j \in D}{p_j}} {|D|} - p_i}
\end{equation}
It is expected to stay below some threshold $comp_{thr}$, otherwise the drones are too far apart, not forming a flock.
 
\noindent\emph{Obstacle clearance}:
Keeping a safe distance to obstacles is required to avoid collisions. We therefore measure the minimum distance from any drone to any obstacle:
\begin{equation}
clear_{obj} = \min_{i \in D; k \in K} \norm{\mathcal{P}(p_i) - p_k}
\end{equation}
For safety, this should always be greater than some threshold $clear_{thr}$.  For obstacles with radius $r_k$, we set $clear_{thr}$ to $r_{\mathit{drone}} + r_{k} + r_{\mathit{safety}}$, where $r_{\mathit{safety}}$ is a safety margin.

\section{Experiments}
\label{sec:eval}

We evaluated SPC using simulation experiments and experiments with actual hardware drones.

\subsection{Simulation Experiments}
\label{sec:sim}

\fullonly{
\begin{figure}[t]
    \centering
     \centering
     \begin{subfigure}[t]{0.31\columnwidth}
         \centering
         \begin{overpic}[width=\textwidth]{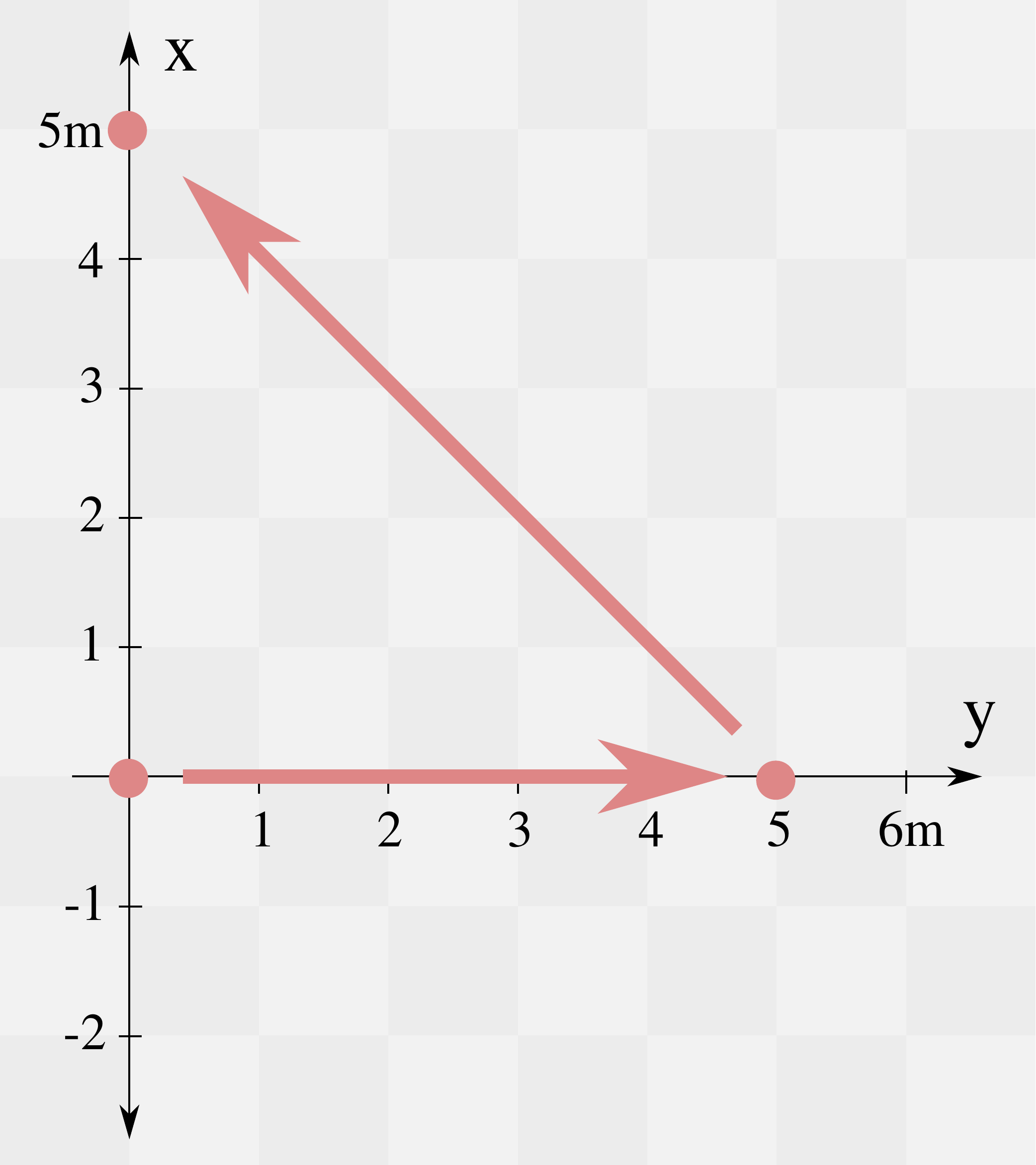}
            \put (5,106) {\textbf{a)}}
         \end{overpic}
        \phantomsubcaption%
        \label{fig:expsim:c}%
     \end{subfigure}
     \hfill
     \begin{subfigure}[t]{0.31\columnwidth}
         \centering
         \begin{overpic}[width=\textwidth]{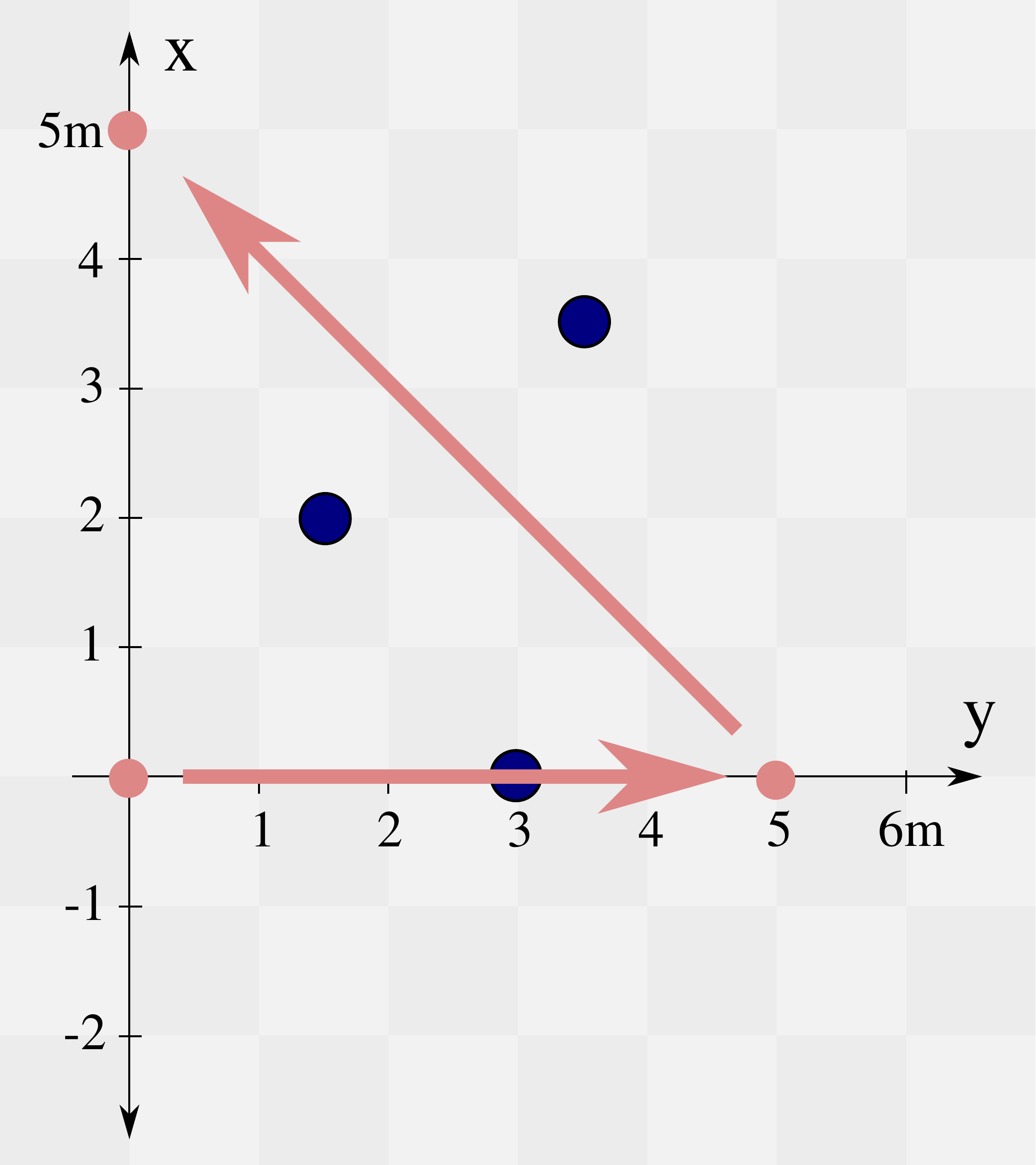}
            \put (5,106) {\textbf{b)}}
         \end{overpic}
        \phantomsubcaption%
        \label{fig:expsim:a}%
     \end{subfigure}
     \hfill
     \begin{subfigure}[t]{0.31\columnwidth}
         \centering
         \begin{overpic}[width=\textwidth]{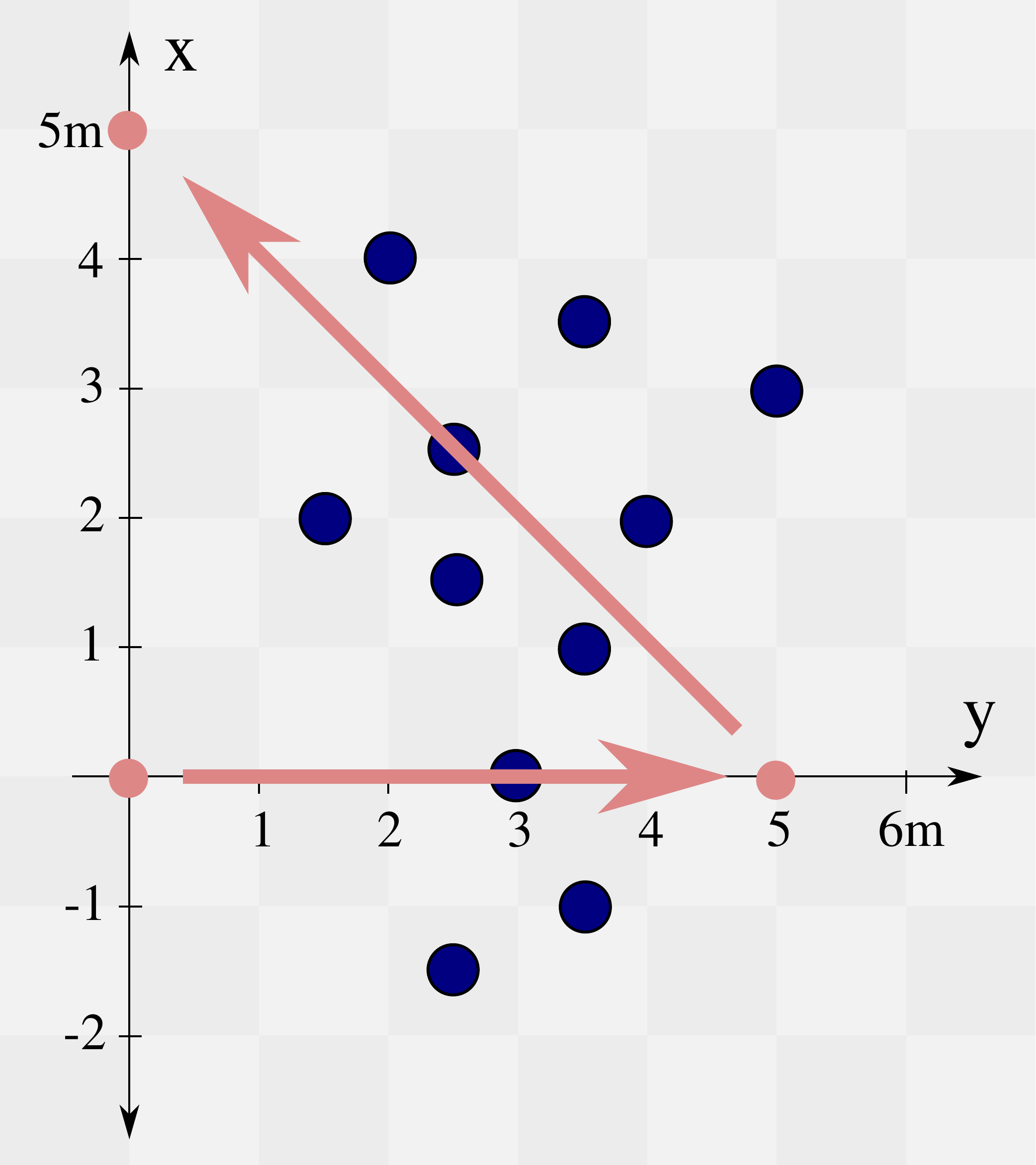}
            \put (5,106) {\textbf{c)}}
         \end{overpic}
        \phantomsubcaption%
        \label{fig:expsim:b}%
     \end{subfigure}\\
     \begin{subfigure}[t]{0.56\columnwidth}
         \centering
         \begin{overpic}[width=\textwidth]{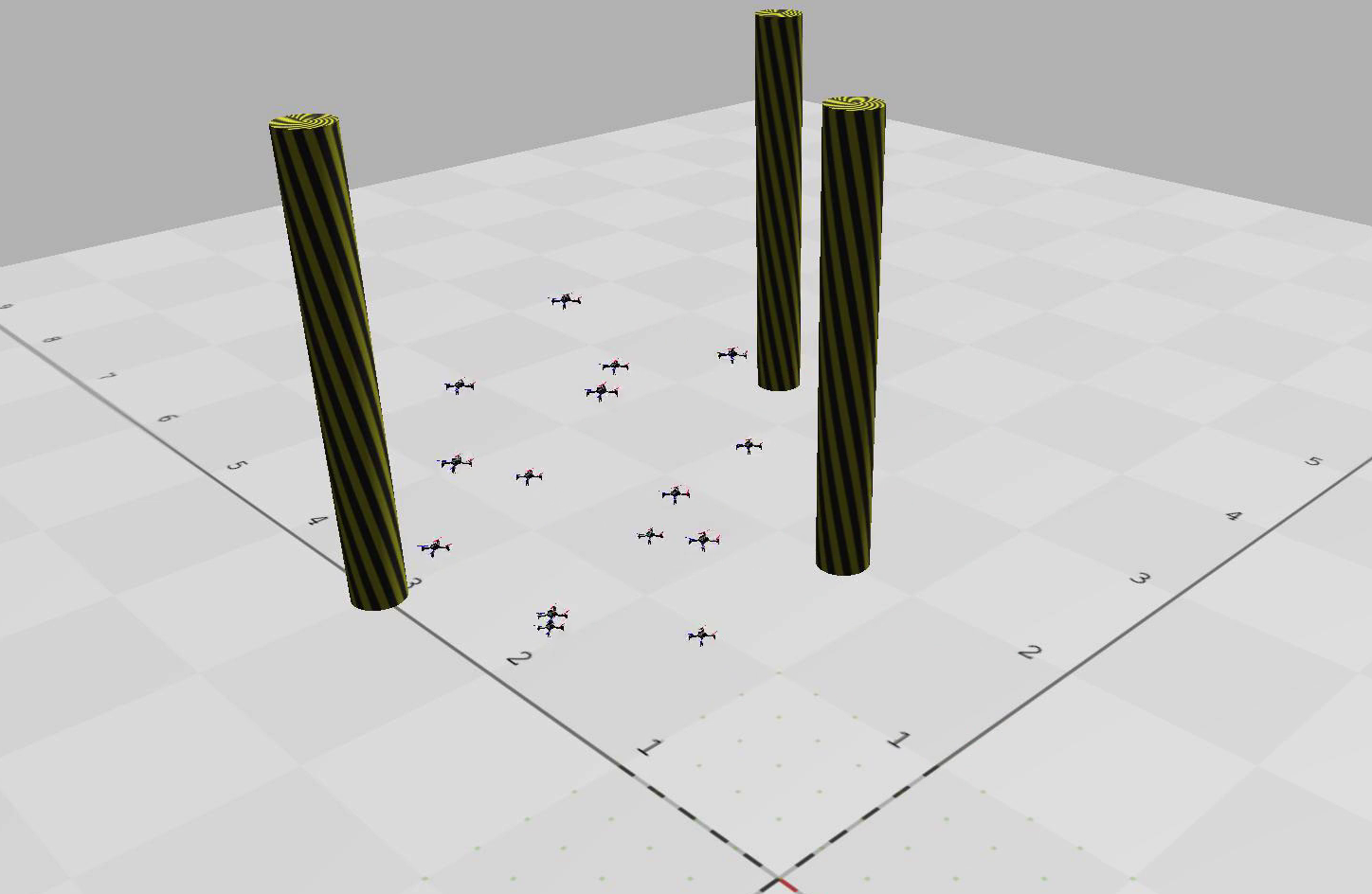}
            \put (-10,60) {\textbf{d)}}
         \end{overpic}
         \phantomsubcaption%
         \label{fig:expsim:screen}%
     \end{subfigure}
     \caption{Simulations were performed using three different scenarios. \textbf{a}: Without obstacles; \textbf{b}: with 3 obstacles; and \textbf{c}: with 11 obstacles indicated in dark-blue. The direct path between the target locations $p_{tar}$ is shown in red. \textbf{d}: Snapshot of a flock with 15 drones in scenario with 3 obstacles.}%
     \label{fig:expsim}%
\end{figure}
}
\shortonly{
\begin{figure}[t]
    \centering
     \centering
     \begin{subfigure}[t]{0.27\columnwidth}
         \centering
         \begin{overpic}[width=\textwidth]{images/drawing_obstacles_and_path_v3a.pdf}
            \put (5,106) {\textbf{a)}}
         \end{overpic}
        \phantomsubcaption%
        \label{fig:expsim:a}%
     \end{subfigure}
     \hfill
     \begin{subfigure}[t]{0.27\columnwidth}
         \centering
         \begin{overpic}[width=\textwidth]{images/drawing_obstacles_and_path_v3b.pdf}
            \put (5,106) {\textbf{b)}}
         \end{overpic}
        \phantomsubcaption%
        \label{fig:expsim:b}%
     \end{subfigure}
     \hfill
     \begin{subfigure}[t]{0.38\columnwidth}
         \centering
         \begin{overpic}[width=\textwidth]{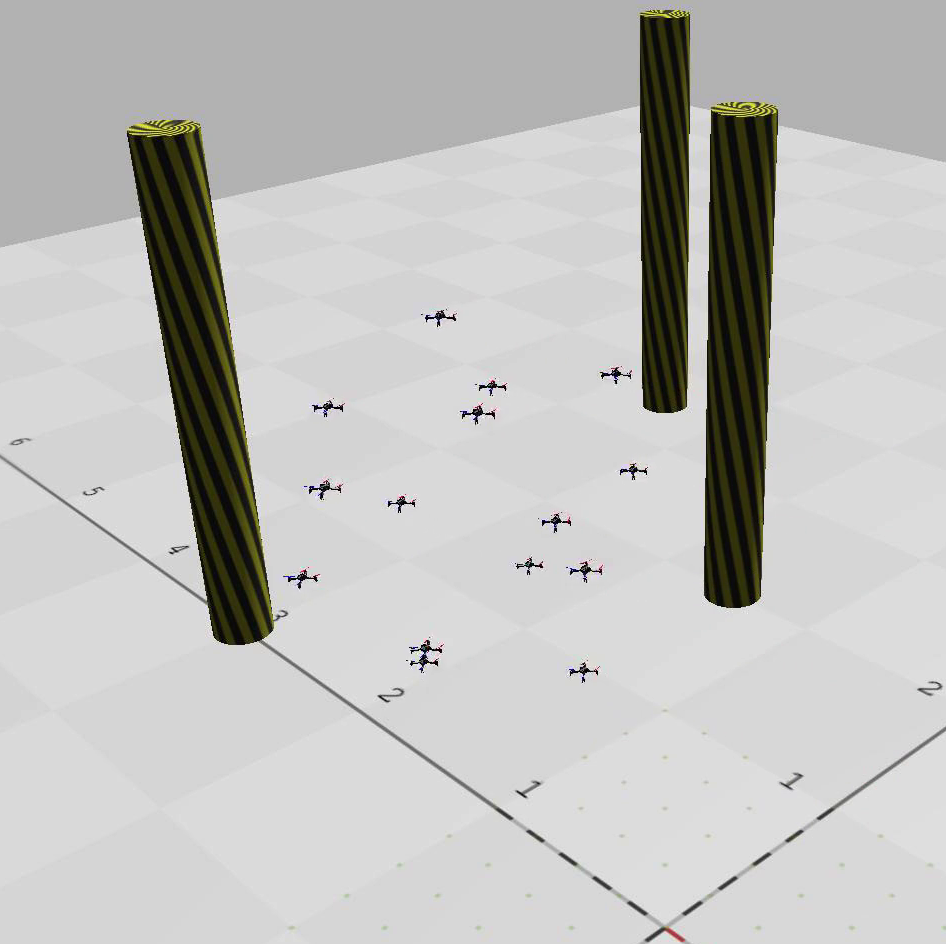}
            \put (-12,88) {\textbf{c)}}
         \end{overpic}
         \phantomsubcaption%
         \label{fig:expsim:screen}%
     \end{subfigure}
     \caption{Simulations were performed using three different scenarios: without obstacles (not shown); \textbf{a}: with 3 obstacles; and \textbf{b}: with 11 obstacles indicated in dark-blue. The direct path between the target locations $p_{tar}$ is shown in red. \textbf{c}: Snapshot of a flock with 15 drones in scenario with 3 obstacles.}%
     \label{fig:expsim}%
\end{figure}
}

\fullonly{
\begin{figure}[t]
    \centering
    \includegraphics[width=0.999\columnwidth]{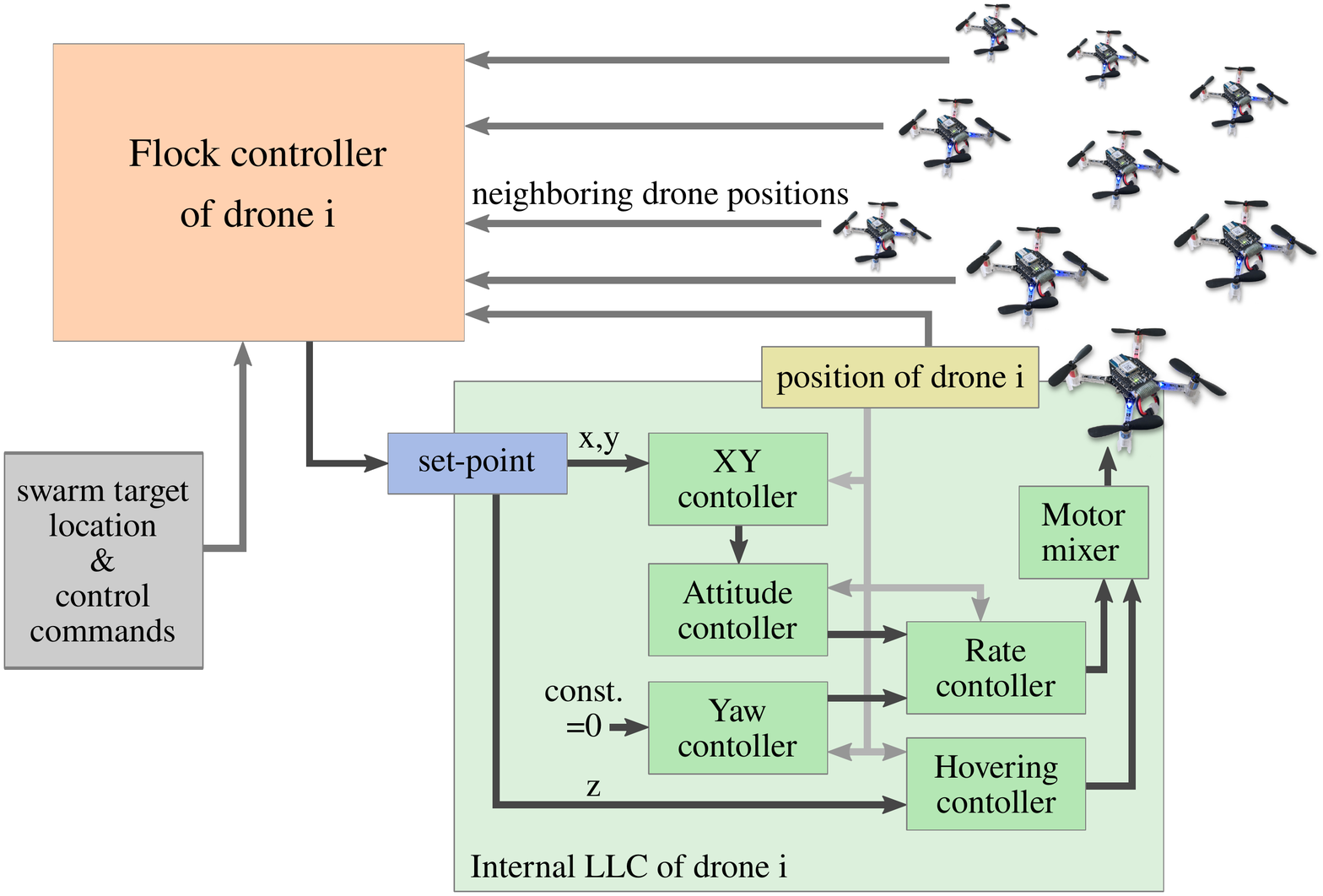}
    \caption{The ROS-node of the SPC controller for a drone $i$ receives position messages of all drones and control messages (e.g. swarm target location). It outputs the set-point for the LLC.}
    \label{fig:swarmcontroller}
\end{figure}
}
\hiddenonly{
\begin{figure}[b]
    \centering
    \includegraphics[width=0.999\columnwidth]{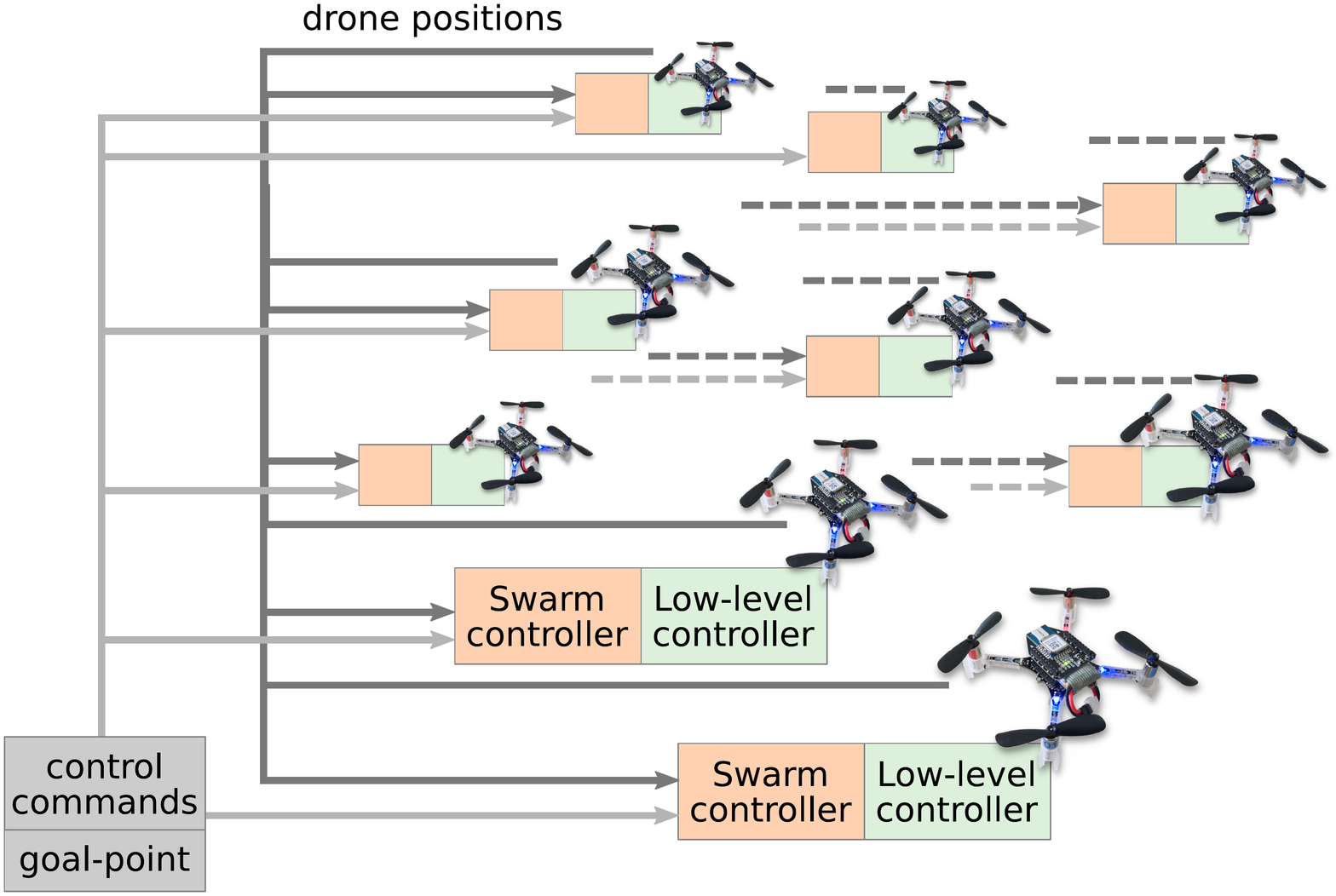}
    \caption{The Swarm controller is fully distributed. There is no further information exchange except from position updates. \todo{is this diagram useful?}}
    \label{fig:swarmdistributed}
\end{figure}
}

As a simulation framework, we use \emph{crazys}~\cite{crazys}, which is based on the \emph{Gazebo} \cite{Gazebo} physics and visualization engine and the Robotic Operating System (ROS) \cite{ros}.
Our SPC algorithm is implemented in C++ as a separate ROS node. 
\fullonly{As shown in Figure~\ref{fig:swarmcontroller}, it}\shortonly{It} receives position messages from neighboring drones, and control messages, such as the target location or a stop command, from the human operator.  It outputs a set-point to the LLC. The SPC we implement is fully distributed: no central optimizer and no further information is exchanged between ROS nodes. 
The SPC node calculates the gradient according to Eqs.~\eqref{eq:grad:coh}-\eqref{eq:grad:obs2}. The spatial look-ahead parameter $N$ is determined dynamically based on the distance to the target location:\vspace{-2mm}
\begin{equation}
N\,{=}\,\ceil*{N^*\,{\cdot}\,max(1,min(1.5\,{\cdot}\,(\norm{p_i\,{-}\,p_{tar}}\,{+}\,0.5), 3))}
\label{eq:dynn}
\end{equation}
This allows the drones to more quickly reach distant target locations and reduces the controller's computational cost (by reducing $|Q_i|$) once the flock reaches the target.
Thereafter, the set-point position $x_{io}^{(r)}$ is determined by Eqs.~\eqref{eq:grad:spat01}-\eqref{eq:grad:spat02}. Auxiliary functions, like hovering at the starting position, 
are also implemented in this node.
In the simulations, we added Gaussian sensor noise, with $\sigma = 10cm$, for drone position measurements.
Cost-function weights and controller parameters (Table~\ref{tab:parameters}) were determined empirically by analysis of the controller behavior.

\begin{table}[h]
\begin{center}
\begin{small}
\begin{tabular}{r|lrrr}
\toprule
~ & ~ & LLC A & LLC B & Hardware\\
\midrule
Cost weights & $\omega_{coh}$ & $20\,m^{-1}$ & $20\,m^{-1}$ & $20\,m^{-1}$ \\
~ & $\omega_{sep}$ & $9\,m^{-1}$ & $9\,m^{-1}$ & $9\,m^{-1}$\\
~ & $\omega_{tar}$ & $150\,m^{-1}$ & $150\,m^{-1}$ & $150\,m^{-1}$ \\
~ & $\omega_{obs}$ & $12\,m^{-1}$ & $12\,m^{-1}$ & n.a.\\
SPC parameters & $N^*$ & $5$ & $3$ & $3$\\
~ & $\epsilon$ & $0.06\,m$ & $0.06\,m$ & $0.025\,m$\\
PFC gain & $k$ & $0.007\,m$ & $0.005\,m$ & n.a. \\
Dimensions & $r_{\mathit{drone}}$ & $0.07\,m$ & $0.07\,m$ & $0.07\,m$ \\
~ & $r_{k}$ & $0.15\,m$ & $0.15\,m$ & n.a. \\
~ & $r_{\mathit{safety}}$ & $0.06\,m$ & $0.06\,m$ & $0.06\,m$ \\
\bottomrule
\end{tabular}
\caption{
Parameters used in simulation experiments and hardware experiments.
}
\label{tab:parameters}
\end{small}
\end{center}
\end{table}

In order to evaluate SPC and its implementation, we defined a path, as shown in Figure~\ref{fig:expsim}. The points of the path are provided in a timed sequence as target location $p_{tar}$. There are three scenarios: without obstacles \fullonly{(Figure~\ref{fig:expsim:c})}, with 3 obstacles (Figure~\ref{fig:expsim:a}), and with 11 obstacles (Figure~\ref{fig:expsim:b}). Simulations were done with flocks of size $|D|\,{=}\,4$, $9$,~$15$,~and $30$.
The neighborhood is defined by $H_i\,{=}\,\{p_j\,{\in}\,D\, {\land}\,\norm{p_i\,{-}\,p_j}\,{<}\,r_H\}$, where $r_H=0.9\,m$.

\fullonly{To check SPC's robustness to different LLCs, we use two LLCs, with a different step response, as shown in Figure~\ref{fig:stepcomparision}.}
\shortonly{To check SPC's robustness to different LLCs, we experimented with two LLCs, with different step responses. LLC~B reaches its set-point for $x$- and $y$-dimensions in less than half the time of LLC~A, while overshooting by about $50\,\%$ more. The LLCs behave very similarly in the z-dimension.}
Figure~\ref{fig:expsim:screen} shows a snapshot of the simulation of a flock with 15 drones. 
\shortonly{A video is provided in the Supplementary Material.}
\fullonly{A video is available at \url{https://youtu.be/ClM2t9eiCsA}.}

\subsubsection{Results}
\begin{figure}[t]
    \centering
     \begin{subfigure}[b]{0.98\columnwidth}
         \centering
         \includegraphics[width=\textwidth]{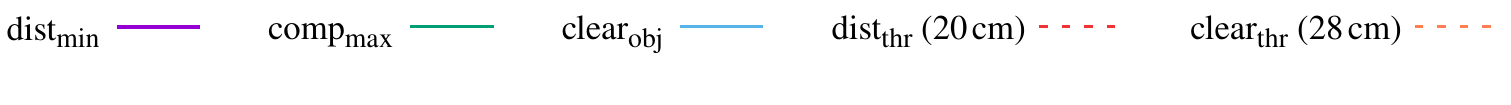}
     \end{subfigure}
      \begin{subfigure}[b]{0.49\columnwidth}
        \centering
        \begin{overpic}[width=\textwidth]{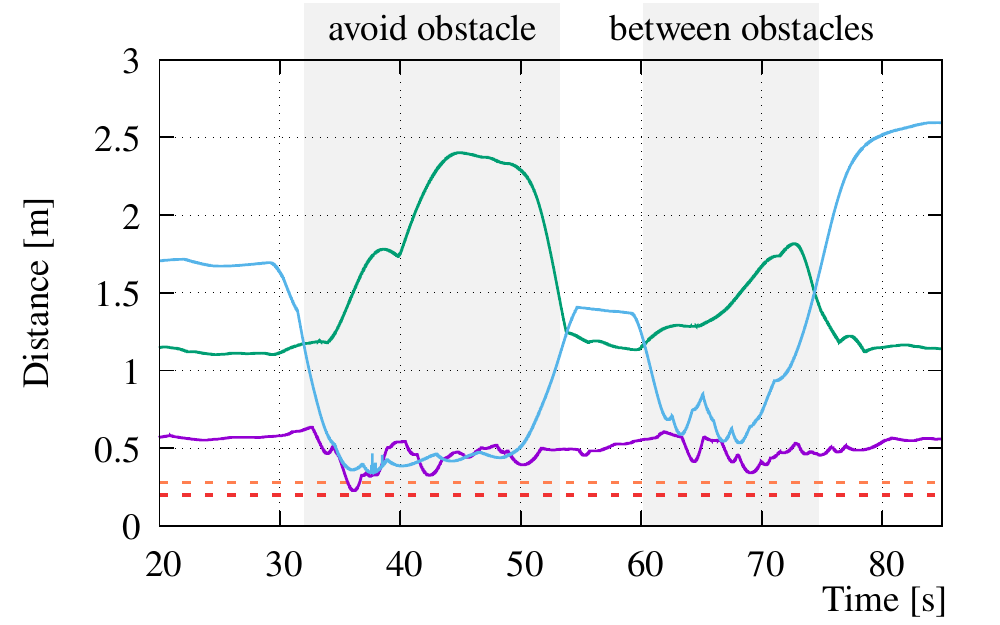}
            \put (15,-2) {{\scriptsize Simulations with LLC A}}
         \end{overpic}
         \vspace*{-1mm}
     \end{subfigure}
     \hfill
     \begin{subfigure}[b]{0.49\columnwidth}
         \centering
        \begin{overpic}[width=\textwidth]{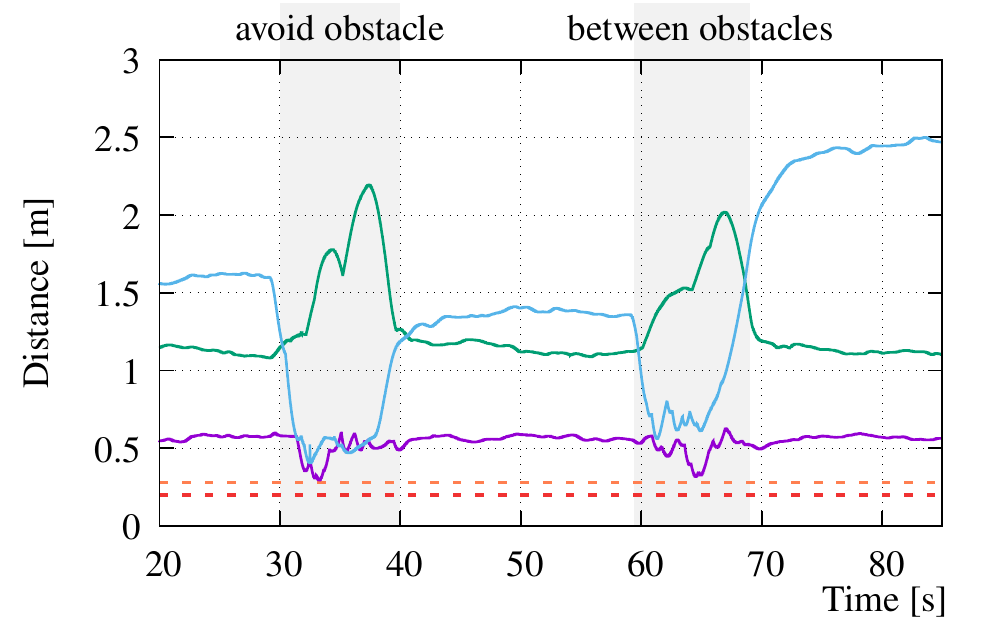}
            \put (15,-2) {{\scriptsize Simulations with LLC B}}
         \end{overpic}
         \vspace*{-1mm}
     \end{subfigure}
    \caption{Quality metrics for SPC simulation with 3 obstacles. Results are aggregated over runs with 4, 9, 15 and 30 drones by using $min$ of $dist_{min}$, $clear_{obj}$ and $max$ of $comp_{max}$.
    }
    \label{fig:sim15plots}
\end{figure}

The analysis of the quality metrics for \emph{collision avoidance}, \emph{compactness}, and \emph{obstacle clearance} show that our control approach successfully maintains a stable flock.
In Figure~\ref{fig:sim15plots}, metrics are plotted over time for one representative simulation for each LLC.
Data from the prefix of an execution, when the drones move from random starting positions into flock formation, are omitted when computing the metrics.
%
In the two shaded time intervals, the flock needs to avoid a single obstacle in the middle of the path, and pass between two obstacles, respectively. It can be observed that some drones are delayed in front of the obstacles, and therefore the \emph{compactness} of the flock temporarily degrades ($comp_{max}$ increases). 

\begin{figure}[t]
    \centering
     \begin{subfigure}[b]{0.98\columnwidth}
         \centering
         \includegraphics[width=\textwidth]{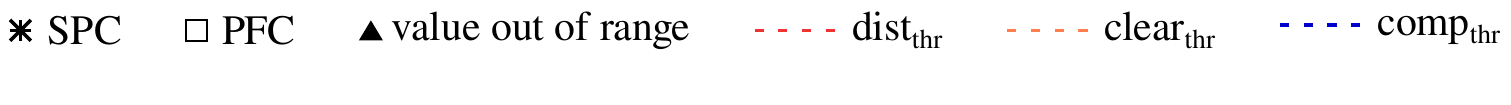}
     \end{subfigure}
    \begin{subfigure}[b]{\columnwidth}
         \centering
         \begin{overpic}[height=53mm]{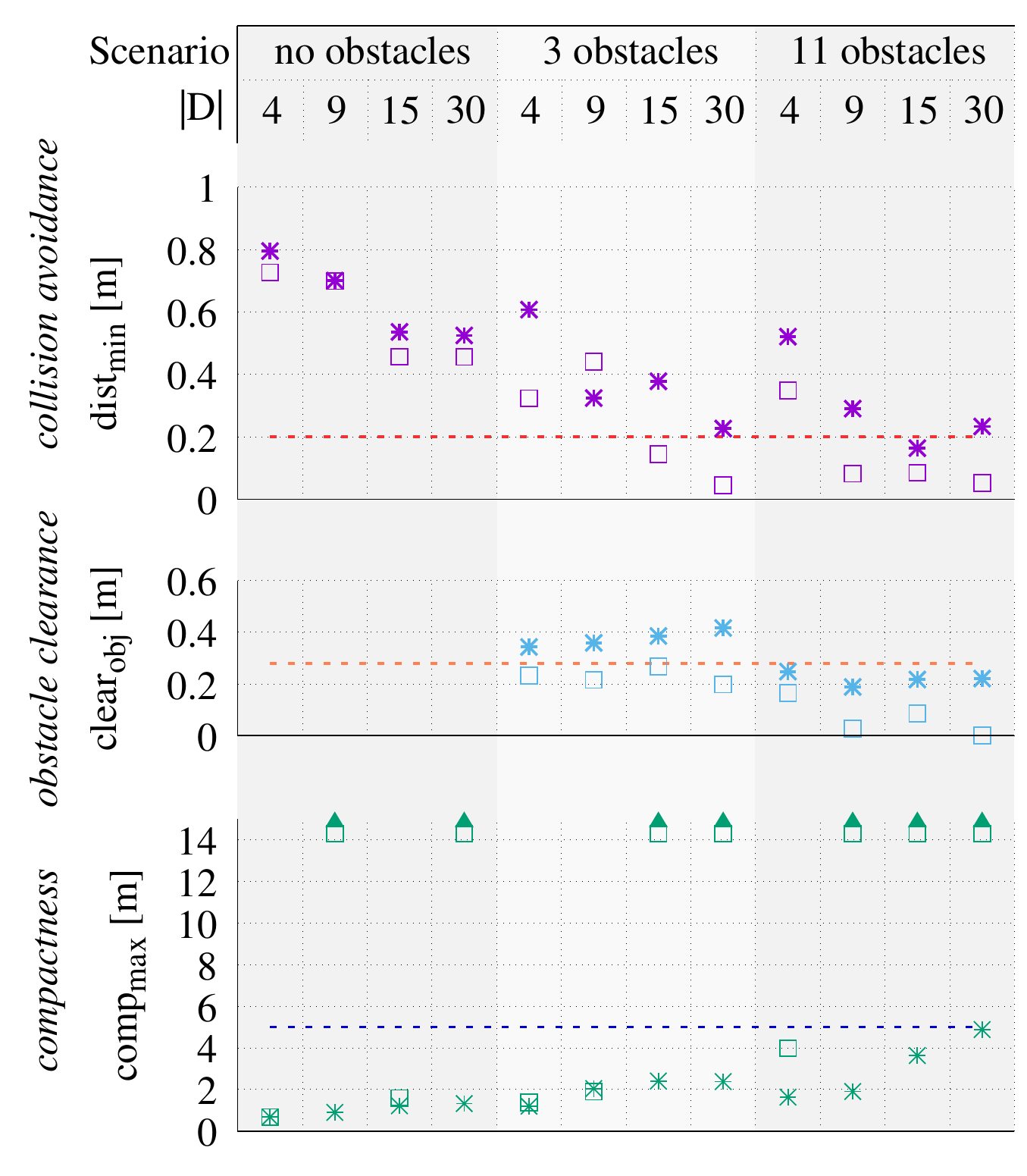}
            \put (22,-4) {{\scriptsize Simulations with LLC A}}
         \end{overpic}
         \begin{overpic}[height=53mm]{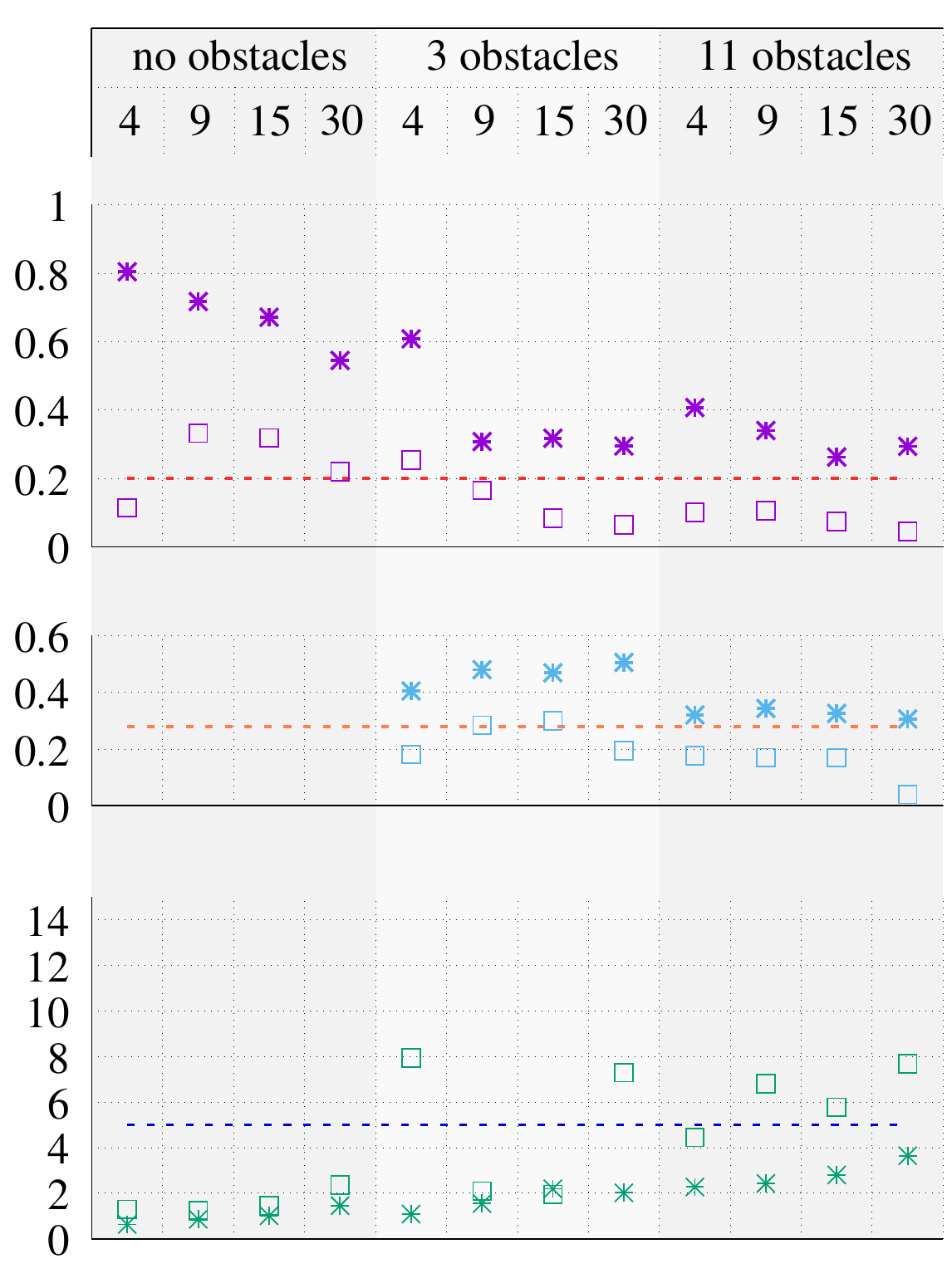}
            \put (10,-4) {{\scriptsize Simulations with LLC B}}
         \end{overpic}
         \vspace*{2mm}
    \end{subfigure}
    \caption{Performance comparison for SPC and PFC. Values are $min$ (for collision avoidance and obstacle clearance) or $max$ (for compactness) over the simulation.
    SPC satisfies $dist_{thr}$ in every scenario but one, while PFC frequenty violates it, especially in the presence of obstacles.
    The scenario with~11 obstacles is intentionally
    cluttered and shows the limits even for SPC
    using LLC~A, as it violates $clear_{thr}$ in this case.
    SPC maintains $comp_{thr}$ in every scenario, while for PFC, $comp_{max}$ sometimes gets very high, even out of range. 
    }
    \label{fig:metrics-new}
\end{figure}

\fullonly{
\begin{figure}[h]
    \centering
     \begin{subfigure}[b]{0.75\columnwidth}
         \centering
         \includegraphics[width=\textwidth]{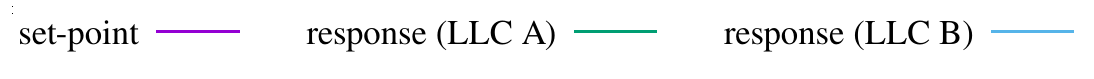}
     \end{subfigure}
     \begin{subfigure}[b]{0.42\columnwidth}
         \centering
         \includegraphics[width=\textwidth]{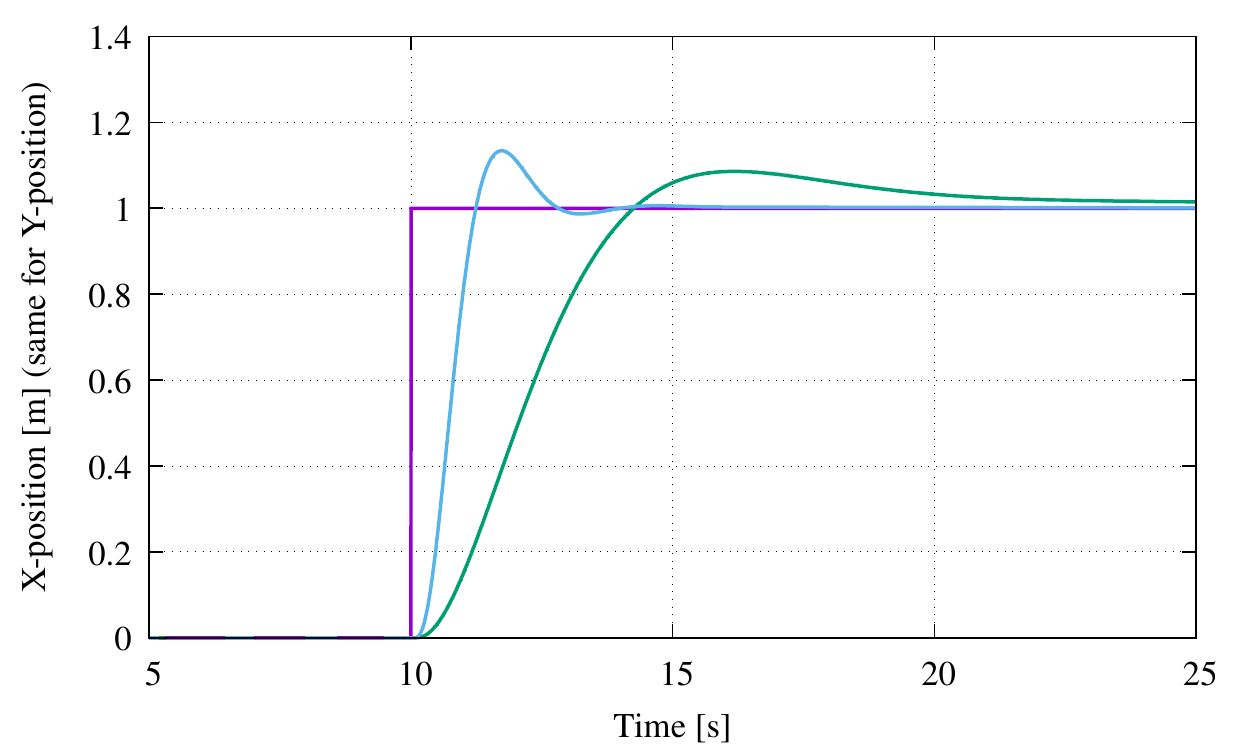}
     \end{subfigure}
     \begin{subfigure}[b]{0.42\columnwidth}
         \centering
         \includegraphics[width=\textwidth]{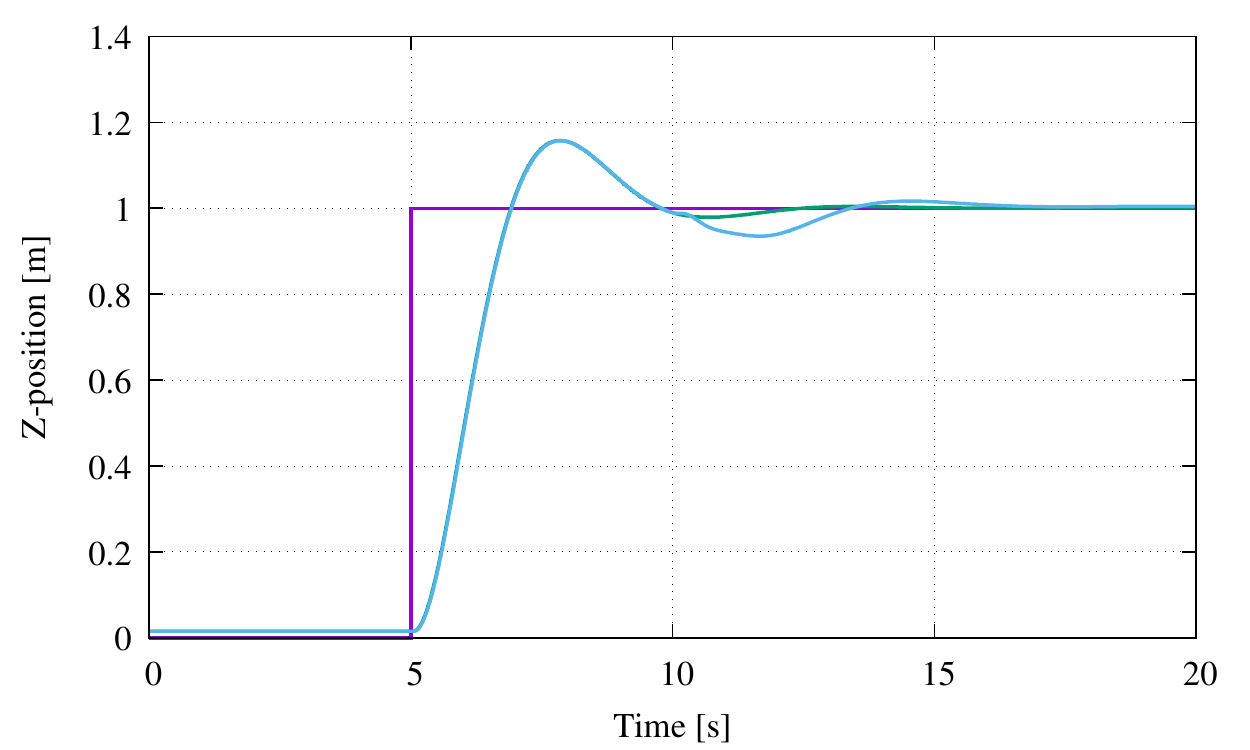}
     \end{subfigure}
    \caption{Step-response for two different LLCs.}
    \label{fig:stepcomparision}
\end{figure}
}

\subsubsection{Comparison with PFC} 

To compare SPC with
~\cite{schwager2009gradient,Tanner03stabilityof}, we also experimented with a \emph{potential-field controller} (PFC) based solely on gradients.  In this controller, the gradient vector $\nabla_{p_i}c$ is used to determine the next set-point $x_{io}^{(r)}$ for the LLC as follows:
\begin{equation}
x_{io}^{(r)} = p_i - k \cdot \nabla_{p_i}c(p_i)
\end{equation}
We determined the gain $k$ empirically such that the target of the flock was reached within the same time as our SPC implementation.
The gain determines how aggressively the controller moves the drone toward the target location.  In the experiments detailed below, the gain is constant, as in~\cite{schwager2009gradient,Tanner03stabilityof}.  We also briefly experimented with dynamic gain, where $k$ is computed using a  function similar to the one in Eq.~\eqref{eq:dynn}.  This had relatively small effects.  Compared to the results with static gain reported below: collision avoidance improved slightly for some scenarios; obstacle clearance improved for some cases with LLC~A, while it worsened with LLC~B; and compactness improved moderately.


Figure~\ref{fig:metrics-new} shows performance metrics for simulations of SPC and PFC controllers.
While both perform reasonably well for LLC~A without obstacles, SPC's performance is superior in the presence of obstacles (only a few violations of $clear_{thr}$). This validates our hypothesis that SPC is particularly valuable when the cost function is more nonlinear (adding obstacles has that effect).
For LLC~B, the PFC controller is hardly able to maintain separation between drones and also from obstacles. Whenever a drone enters or leaves  another drone's neighborhood, the cost function instantaneously changes its value; the gradient changes too. This causes the PFC controller to fail: in these simulations, we observed oscillating behavior and multiple collisions. SPC successfully deals with all of these situations. {\em{In short, SPC is more robust to nonlinearities in the cost function and differences in the behavior of the LLC.}}

\subsection{Hardware Experiments}

We also experimented with real drones, specifically, \emph{Crazyflie 2.1}-quadcopters~\cite{Crazyflie}; see Figure~\ref{fig:exp9:crazyflie}. For localization, we used the \emph{Loco-Positioning} system~\cite{lps}.
The drones seamlessly integrated with the localization system, resulting in a (internally) positional LLC that enables a drone to hold its position at a given set-point.  Stability, however, depends on both the accuracy of the localization system and on the mechanical limitations of the drone. 
When hovering at a given set-point, we observed noise in the drone's position in the range of 15\,cm.
This was also noted in~\cite{uwbaccuracy}.

In the hardware implementation, we used ROS with the same software node as in Section~\ref{sec:sim}, with only minor parameter modifications. This demonstrates the robustness of SPC with respect to a potential sim-to-real transfer gap. Since Crazyflies are incapable of running ROS on-board, we transmit the position updates to a PC that runs the controller and transmits the set-point position to the drone.
Our experiments therefore also show that SPC is resilient to the additional delay introduced by radio transmission of position updates and set-point messages.
Our controller, however, could be ported to run directly on ROS-capable drones, since we run it separately for each drone. 

\subsubsection{Results}

Due to the limited space available in our robotics lab, we could fly only simple trajectories with a flock of nine Crazyflies. We were, however, able to show that SPC successfully maintains a stable flock while hovering and while moving between target locations.  Figure~\ref{fig:exp9:vid} shows an image of our flock. \shortonly{A video is provided in the Supplementary Materials.}
\fullonly{A video is available at \url{https://youtu.be/ClM2t9eiCsA}.}


Quality metrics for \emph{collision avoidance} and \emph{compactness} are given in Figure~\ref{fig:exp9qual}. The jaggedness of the plots is due to the sensing noise introduced by the localization system. Nevertheless, the plots show that our SPC controller continuously maintained a safe distance between drones and achieved a compact flock formation.

\begin{figure}[t]
    \centering
     \centering
     \begin{subfigure}[c]{0.17\columnwidth}
         \centering
         \begin{overpic}[width=\textwidth]{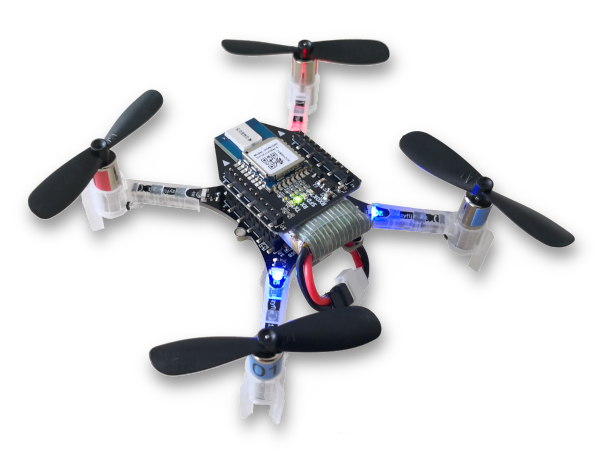}
            \put (0,95) {\textbf{a)}}
         \end{overpic}
         \phantomsubcaption
         \label{fig:exp9:crazyflie}
     \end{subfigure}
     \begin{subfigure}[c]{0.31\columnwidth}
         \centering
         \begin{overpic}[width=\textwidth]{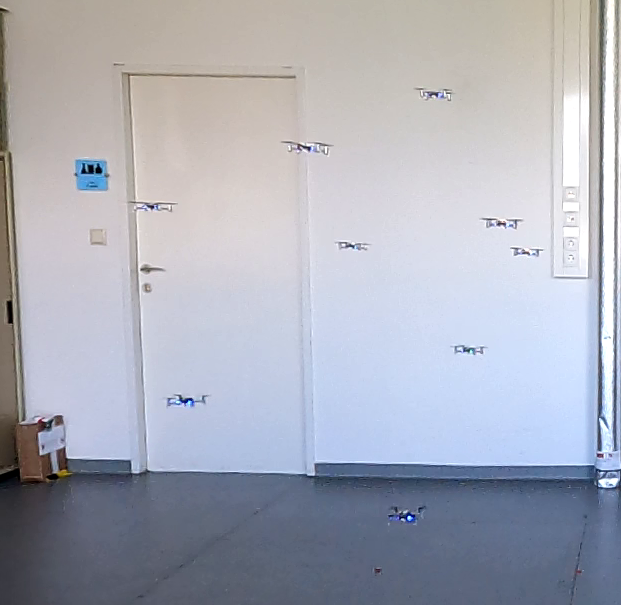}
            \put (-15,80) {\textbf{b)}}
         \end{overpic}
         \phantomsubcaption
         \label{fig:exp9:vid}
     \end{subfigure}
     \hfill
     \begin{subfigure}[c]{0.49\columnwidth}
         \centering
         \begin{overpic}[width=\textwidth]{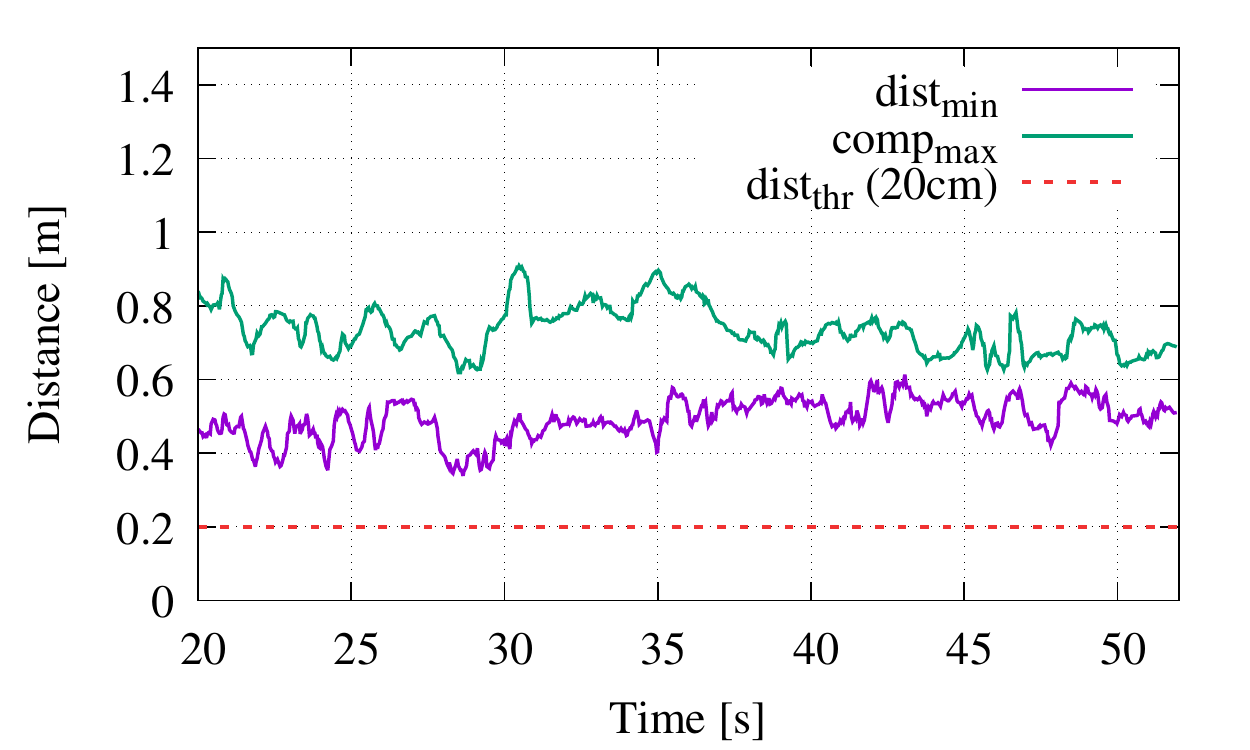}
            \put (0,50) {\textbf{c)}}
         \end{overpic}
         \phantomsubcaption
         \label{fig:exp9qual}
     \end{subfigure}
    \caption{Hardware experiments. \textbf{a}: Crazyflie quadcopters were used. \textbf{b}: A flock of 9 drones in our robotics lab; \shortonly{a video is in the Supplementary Materials}\fullonly{a video is available at https://youtu.be/ClM2t9eiCsA}. \textbf{c}: Quality metrics.}
    \label{fig:exp9}
\end{figure}

\section{RELATED WORK}
\label{sec:related}
Reynolds~\cite{Reynolds1987} was the first to propose a flocking model,  using cohesion, separation, and velocity alignment force terms to compute agent accelerations.
Reynolds model was extensively studied~\cite{Eversham2010} and adapted for different application areas~\cite{SoonJo2018}.
Alternative flocking models are considered in \cite{Olfati-Saber}, \cite{Declarative2018}, \cite{Samuel2014}, \cite{Soria2019}, and \cite{Tanner03stabilityof}. In these approaches, flocks are described using point models. This means that physical properties of agents (e.g., drones) such as mass and inertia, are not taken into account.
In our work, we evaluate SPC on a 
highly realistic physical drone model, as well as on real hardware.

In addition to these largely theoretical approaches, in~\cite{Vrhelyieaat2018} and~\cite{Soria2021}, flocking controllers are implemented and tested on real hardware. However, the approach of~\cite{Soria2021} involves the use of nonlinear model-predictive control (NMPC), which is computationally more expensive than SPC. In contrast to SPC, \cite{Vrhelyieaat2018} requires the velocity of neighboring drones.
Gradient optimization for robot control has been studied in~\cite{schwager2009gradient}. In contrast, SPC uses spatial look-ahead as opposed to pure gradient descent.

\section{CONCLUSIONS}
\label{sec:conclusion}

We introduced the novel concept of \emph{Spatial Predictive Control} (SPC), and demonstrated its utility on the drone flocking problem.
%
We performed simulation experiments using a physics engine with a detailed drone model.  Our results demonstrated SPC's ability to form and maintain a flock, avoid obstacles, and move the flock to multiple target locations.  They also highlighted SPC's robustness to sensing noise and LLC variability, and its role in the controller hierarchy.

We also evaluated the same controller implementation on a flock of nine Crazyflie~2.1 quadcopters, thereby demonstrating the effectiveness of SPC in controlling real hardware.
Needing only a minor parameter adjustment, and no modifications to the control algorithm, SPC proved to be very robust in terms of a potential sim-to-real transfer gap.  The hardware demonstration also highlighted SPC's capability to perform properly in the presence of significant sensor noise introduced by the localization system and the extra latency introduced by radio transmission of position and control signals.

We also experimentally compared SPC with the closely-related PFC-based approach of~\cite{schwager2009gradient}.  We found that SPC exhibits superior performance and stability, as its discrete search for an optimal solution enables it to avoid oscillations. SPC copes significantly better with nonlinear cost functions, and is more resilient in general. SPC is a general technique for designing {\em{middle-level controllers}} sandwiched between high-level planners and LLCs that often come integrated with the hardware.
As future work, we plan to evaluate SPC on other robotic control \& planning problems, 
and to conduct a demonstration of SPC in (outdoor) settings that lack an absolute localization system (GPS-denied environments).

\fullonly{
\section*{ACKNOWLEDGMENTS}

R.G. was partially supported by EU-H2020 Adaptness and AT-BMBWF DK-RES and CPS/IoT Ecosystem. This work was supported in part by NSF grants %
CCF-1954837, 
CCF-1918225, 
and CPS-1446832 
and ONR grant N000142012751. 
}


\fullonly{\clearpage}
\renewcommand*{\bibfont}{\footnotesize}
\printbibliography 

\end{document}